\documentclass[conference]{IEEEtran}
\IEEEoverridecommandlockouts
\usepackage{cite}
\usepackage{amsmath,amssymb,amsfonts}
\usepackage{algorithmic}
\usepackage{graphicx}
\usepackage{textcomp}
\usepackage{xcolor}
\usepackage{subfig}
\usepackage{todonotes}
\usepackage{hyperref}
\usepackage{fancyvrb}
\usepackage{verbatimbox}

\usepackage{todonotes}

\def\BibTeX{{\rm B\kern-.05em{\sc i\kern-.025em b}\kern-.08em
    T\kern-.1667em\lower.7ex\hbox{E}\kern-.125emX}}
\begin{document}

\title{Timeline-based Process Discovery
\thanks{The research by Jan Mendling was supported by the Einstein Foundation Berlin under grant EPP-2019-524 and by the German Federal Ministry of Education and Research under grant 16DII133.}
}

\author{\IEEEauthorblockN{Harleen Kaur, Jan Mendling, Christoffer Rubensson}
\IEEEauthorblockA{
\textit{Humboldt-Universität zu Berlin}\\
Berlin, Germany \\
\{kaurharl,jan.mendling,christoffer.rubensson\}@hu-berlin.de} 
\and
\IEEEauthorblockN{Timotheus Kampik}
\IEEEauthorblockA{
\textit{SAP Signavio}\\
Berlin, Germany \\
timotheus.kampik@sap.com}
}

\maketitle

\begin{abstract}
A key concern of automatic process discovery is to provide insights into performance aspects of business processes. Waiting times are of particular importance in this context. For that reason, it is surprising that current techniques for automatic process discovery generate directly-follows graphs and comparable process models, but often miss the opportunity to explicitly represent the time axis. In this paper, we present an approach for automatically constructing process models that explicitly align with a time axis. We exemplify our approach for directly-follows graphs. Our evaluation using two BPIC datasets and a proprietary dataset highlight the benefits of this representation in comparison to standard layout techniques.\end{abstract}


\section{Introduction}
Process mining techniques serve the purpose of providing insights into process performance, which can help organizations to reduce bottlenecks, waiting times, and operational costs. To this end, process mining research has developed a plethora of algorithms for automatic process discovery~\cite{DBLP:journals/is/WeerdtBVB12}. Many of these algorithms have been designed with the ambition to improve the accuracy of the generated models in terms of metrics like precision and recall~\cite{polyvyanyy2020monotone}. A recent survey identifies 344 papers that are directly concerned with the development of new discovery algorithms and their refinement~\cite{DBLP:journals/tkde/AugustoCDRMMMS19}. These algorithms generate process models using various formalisms, including Petri nets, Declare rules, Process trees, BPMN, and Directly-follows graphs (DFGs)~\cite{DBLP:journals/tkde/AugustoCDRMMMS19}.

A key concern of process performance analysis is the understanding of why and to which extent waiting times occur during the execution of the process. Waiting represents a source of waste and is connected with operational bottlenecks, high inventory costs, and low customer satisfaction~\cite[p.222]{DBLP:books/sp/DumasRMR18}. Against this backdrop, it is remarkable that process mining algorithms by and large focus on the temporal order of events, but almost entirely abstract from the actual time axis~\cite{yeshchenko2022survey}. Some exceptions are techniques for color highlighting of long durations~\cite[p.222]{DBLP:books/sp/DumasRMR18}, performance spectra \cite{denisov2019predictive} inspired by Marey charts~\cite[p.260]{marey1875methode}, or dotted charts~\cite{song2007supporting} that plot event sequences aligned to a time axis. The latter belong to timeline-based analysis techniques such as Gantt charts~\cite{DBLP:series/hci/AignerMST11} and early process charts~\cite{mendling-nordsieck} that have been used for process analysis already in the 1920s. Up until now, there are no techniques that integrate the strengths of timeline-based charts with process models automatically constructed by process mining algorithms.

This paper presents an approach for automatically constructing process models that explicitly align with a time axis. We address this research challenge at the level of graph layout and position activities according to their relative temporal distance to the start of the process. We exemplify this idea for directly-follows graphs, though it can be equally adopted for process mining algorithms that construct other types of graphs like Petri nets or BPMN. We demonstrate the effectiveness of our approach based on a comparison of standard layout with our timeline-based layout.

The remainder of the paper is structured as follows. Section~\ref{back} summarizes prior research on timelines in visual analytics. Section~\ref{approach} presents the concepts of our approach to layout graphs using a timeline. Section~\ref{eval} evaluates the effectiveness of our approach and discusses its implications. Section~\ref{conclusion} concludes with a summary and outlook on future research.

\section{Background}\label{back}
In this section, we discuss prior research on the analysis of event sequence data that explicitly considers a time axis. First, we discuss broadly timeline-based techniques from visual analytics. Then, we highlight the few contributions from process mining research that integrate a time axis.

\subsection{Timelines in Visual Analytics}
Timeline-based visualizations are techniques developed in the area of visual analytics that order event sequence data along a time-axis \cite{DBLP:journals/corr/abs-2006-14291}. Commonly, events represent visual elements on a canvas, such as glyphs in different colors, shapes, and sizes \cite[p. 4]{DBLP:journals/corr/abs-2006-14291}. The timeline represents but is not restricted to a horizontal axis with time indicated as discrete time points (e.g., specific dates) or reference time units (e.g., weeks, days, hours). Recent surveys developed a taxonomy for the classifications of event sequence data visualizations~\cite{yeshchenko2022survey,DBLP:journals/corr/abs-2006-14291}. Regarding timeline-based visualization techniques, the authors define five categories: fixed, duration, converging-diverging, evolution, and combinations of these techniques.

Fixed timeline-based visualization techniques \cite{yeshchenko2022survey} arrange events chronologically, with each event attached to a specific point in time on the timeline. Mostly, the events are depicted as glyphs in circular \cite{DBLP:journals/tvcg/ChenXR18, Kwon2020DPVisVA}, triangular \cite{DBLP:journals/tvcg/MonroeLLPS13, DBLP:journals/cgf/LeiteGMGK20}, or quadrilateral \cite{DBLP:journals/cgf/DortmontEW19} shapes, with colors applied to denote event types \cite{DBLP:journals/tvcg/ChenXR18, DBLP:journals/cgf/DortmontEW19, Kwon2020DPVisVA}. Event sequences are separated into rows stacked along the opposite axis of the timeline \cite{DBLP:journals/tvcg/ChenXR18, Kwon2020DPVisVA}. Although authors commonly follow this standard visual constellation, some deviate by using various visual elements to convey additional information. One example is the use of rectangles by Leite \emph{et al.} \cite{DBLP:journals/cgf/LeiteGMGK20} to highlight specific time points, along with triangles of various sizes, shapes, and positions to represent event types or numerical values. Here, the colored borders of the triangles serve as markers for exceeding a threshold level \cite[p. 350]{DBLP:journals/cgf/LeiteGMGK20}.


Duration timeline-based visualization techniques \cite{yeshchenko2022survey} organize events chronologically along a time axis. They include information about event duration by extending its visual elements across multiple time points. More specifically, rectangular glyphs represent events while simulating event duration by stretching the glyphs along the time axis, e.g., \cite{DBLP:journals/tvcg/VrotsouJC09, DBLP:journals/ivs/VrotsouYC14, DBLP:journals/cgf/RosenthalPMO13, DBLP:journals/cgf/HanRDAS15, DBLP:journals/tvcg/NguyenTAATZ19}. Like fixed timeline-based visualizations, event glyphs are often color-coded to distinguish between different event types \cite{DBLP:journals/tvcg/VrotsouN19, DBLP:journals/tvcg/NguyenTAATZ19, DBLP:journals/tvcg/ZengWWWLEQ20}, emphasize particular phenomena in the data \cite{DBLP:journals/cgf/RosenthalPMO13}, — or both \cite{DBLP:journals/cgf/HanRDAS15}. Han \emph{et al.} \cite{DBLP:journals/cgf/HanRDAS15} deviate somewhat from the standard visual setup by including an overlapping design of event boxes to represent temporal overlappings of behavioral patterns in the data. In another approach, Rosenthal \emph{et al.} \cite{DBLP:journals/cgf/RosenthalPMO13} display duration as lines that extend out boxes that represent problems in airline operations. Varying box sizes are further utilized for encoding information on hierarchical structures, whereas text replaces color as an event-type indicator \cite[p. 85]{DBLP:journals/cgf/RosenthalPMO13}.


Converging-diverging timeline-based visualization techniques \cite{yeshchenko2022survey} use lines on a timeline with altering positions in the opposed direction to show changes in affiliations between event sequences. Each line represents an entity, whereas bundles of lines indicate relational closeness between these entities. As an example of this, Liu \emph{et al.} \cite{DBLP:journals/tvcg/LiuWWLL13} let lines represent a character in a story, or a similar story-related object, with bundled lines indicating relational interactions. Here, lines that move together indicate a start of a new social interaction within the story, whereas the opposite is true for diverging lines \cite[p. 2438]{DBLP:journals/tvcg/LiuWWLL13}. Moreover, this concept of affiliative development is also used to study the spread of diseases \cite{DBLP:journals/corr/abs-2008-09552} or changes in community structures \cite{DBLP:journals/cgf/RedaTJLB11}. Some studies use converging (diverging) lines to emphasize temporal discrepancies \cite{DBLP:journals/tvcg/XuMR017} or mark event changes (state changes) \cite{Kwon2020DPVisVA}. Colors are typically used to differentiate entire event sequences \cite{DBLP:journals/cgf/RedaTJLB11, DBLP:journals/tvcg/LiuWWLL13}, to accentuate temporal outliers \cite{DBLP:journals/tvcg/XuMR017}, or to highlight various sections within the lines themselves \cite{DBLP:journals/corr/abs-2008-09552}. Some authors complement the lines with highlighted background colors to provide additional details \cite{DBLP:journals/tvcg/LiuWWLL13, DBLP:journals/corr/abs-2008-09552}, e.g., to indicate geographical location of the entities \cite[p. 714]{DBLP:journals/corr/abs-2008-09552}.


Evolution timeline-based visualization techniques \cite{yeshchenko2022survey} condense event sequence information into density charts or evolution line graphs and map them along the time axis. These techniques help to analyze the concentration of data and how it develops over time. Wu \emph{et al.} \cite{DBLP:journals/tvcg/WuLYLW14} aggregate event data to show the propagation of opinions on social media (Twitter tweets), using a sequential color map to indicate the degree of positivity or negativity of the aggregated opinions. Sung \emph{et al.} \cite{DBLP:journals/cgf/SungHSCLW17} also use a similar visualization to analyze online lecture satisfaction through video comments, using different colors to indicate event types. Finally, Dortmont \emph{et al.} \cite{DBLP:journals/cgf/DortmontEW19} incorporate a density plot chart into their visual system ChronoCorrelator to provide a global view of the event data.


The final category of \emph{combination} techniques refers to hybrid solutions that combine features of the timeline-based visualization types described above \cite[p. 26]{yeshchenko2022survey}. One stream of approaches combines fixed with duration timeline-based visualizations. This involves using a direct combination of glyphs for single events and bars in variable lengths to enclose a respective time span \cite{DBLP:journals/tvcg/MonroeLLPS13, DBLP:journals/tvcg/FuldaBM16}. In an approach that seamlessly integrates the visual features of both categories, Sung \emph{et al.} \cite{DBLP:journals/cgf/SungHSCLW17} involve grouping events into boxes of various sizes, which are then placed on the time axis to denote average time values. Lines from the boxes are linked to the time axis to indicate the precise time point of each event \cite[p. 149]{DBLP:journals/cgf/SungHSCLW17}. 


Furthermore, another stream of literature on hybrid timeline-based visualizations combines duration with evolution timeline-based visualization techniques. LiveGantt, proposed by Jo \emph{et al.} \cite{DBLP:journals/tvcg/JoHPKS14}, uses a Gantt chart enhanced with an event reordering-aggregation mechanism to compress complex event data into visually comprehensible units. CoreFlow, proposed by Liu \emph{et al.} \cite{DBLP:journals/cgf/LiuKDGHW17}, combines an icicle plot with a node-link visualization to create tree-link structures of the event data. Here, the node-link visualization temporally orders the event sequences along a vertical time axis to display the average time elapsed, whereas the icicle plot partitions the data into hierarchically structured density blocks \cite[pp. 530-531]{DBLP:journals/cgf/LiuKDGHW17}.


\subsection{Process Mining with Time Axis}

In process mining, only a few works integrate a time axis into their approaches. 

Bose \emph{et al.} \cite{DBLP:journals/is/BoseA12} offer a fixed timeline-based visualization for aligning traces along a horizontal time axis. This visualization also called a dotted chart, is similar to a Gantt chart. It uses circular glyphs for color-coded events to indicate their type without indicating activity duration. The events are arranged horizontally along the timeline according to their specific time point (timestamp). Leoni \emph{et al.} \cite{DBLP:journals/dss/LeoniAAH12} offer a fixed timeline-based approach to visualize events (or work tasks) as dots in a timeline map. Colors are used to convey various distance metrics, such as the geographical distance between a resource and a task. Events close to each other are grouped into larger glyphs, whereas the time points on the timeline mark a designated future for events rather than their time of execution.

As for duration timeline-based approaches in process mining, Gantt charts have been used to detect temporal anomalies in event logs. Low \emph{et al.} \cite{DBLP:journals/is/LowAHWW17} suggest a visualization to compare various process-related characteristics of two event logs simultaneously. In a time-based view, bar lengths may represent resource performance, where activity characteristics can be mapped to different color codes. Richter \emph{et al.} \cite{DBLP:journals/is/RichterS19} use bar lengths to encode transition times between activities, displayed horizontally on the time axis in linear time. Each row on the vertical axis represents a transition between two activities, and colors convey temporal trends.

In works by Suriadi \emph{et al.} \cite{DBLP:journals/dss/SuriadiOAH15, DBLP:journals/dss/SuriadiWXAH17}, two evolution timeline-based visualization tools show event properties as color-coded evolution line graphs along a horizontal time axis. The visual elements are similar in both tools yet differ in their analytical focus. Whereas \cite{DBLP:journals/dss/SuriadiOAH15} focuses on event intervals, \cite{DBLP:journals/dss/SuriadiWXAH17} analyzes resource behavior patterns. 


\subsection{Summary and Research Objective}
From this discussion of the literature, we observe contributions on timelines largely abstract from replay semantics as provided by process models. In turn, most process mining techniques do not explicitly map to a time axis, and if they do, they abstract from model semantics. Against this backdrop, we define the research objective of our work to develop an approach for automatically constructing process models that explicitly align with a time axis.

\section{Timeline-based Discovery}\label{approach}
In this section, we present the concepts of our approach to transform an event log to represent relative physical time and to augment the traditional DFG visualization with a timeline representing mean relative occurrence time of events.

\begin{figure*}[ht]
    \includegraphics[width=\textwidth]{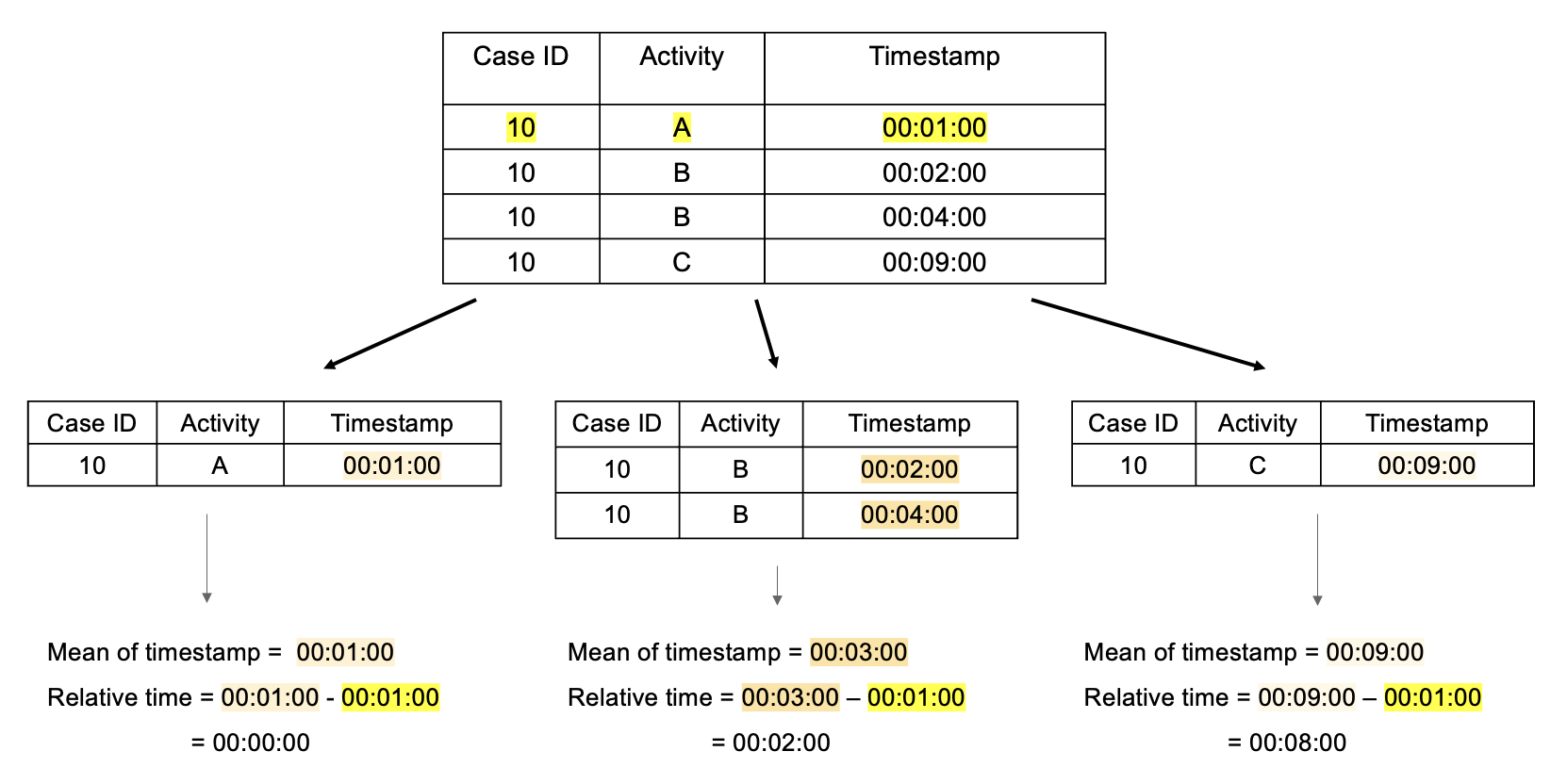}
    \caption{Diagram showing how the relative time is calculated for each case in an event log. First, the event log is sorted according to the case ID and timestamp. For each case, the earliest activity is taken to be the start activity, which is activity $A$ in this diagram, highlighted in yellow. Case ID $10$ has a loop as activity $B$ is being repeated within a case. Each case is then split into a sub-table for each activity to deal with loops. A mean value is calculated for each sub-table's timestamp column. The relative time is calculated by subtracting the initial timestamp value from the mean of the value of all timestamps for a given activity.}
    \label{fig:diagram}
\end{figure*}
%

\subsection{Preliminaries}

We assume an event log $L$ is a finite set of traces and every trace $C \in L$ is a finite set of event tuples such that for every $(a, t) \in C$ it holds that $a \in A$, where $A$ is a finite background set of activity identifiers and $t \in \mathbb{N}_0$\footnote{Here, we use $\mathbb{N}_0$ as an approximate domain for Unix time. Note that a uniqueness assumption for timestamps can typically be made (or enforced); hence, assuming a multiset of traces or event tuples is not necessary.}.
To visualize the event log from the perspective of relative physical time, we need to adjust the timestamps in the event log so that they are relative to the start time of the case and not logged in global time (Figure~\ref{fig:diagram}):

\begin{enumerate}
    \item  \textbf{Determine case start time.} For each case, we determine the case start time, \emph{i.e.}, the minimal time for which any event is logged in the case: we define $t_{min}(C) := avg(\{t | (a, t) \in C, \forall (a', t') \in C: t \leq t' \})$ ($avg$ is used as an aggregation function for merely technical purposes).
    \item \textbf{Determine relative timestamps.} For each case, for each event, we determine the relative (to the case's start time) timestamp of the event by substracting the case start time from the timestamp of the event: we can introduce a function $f$ 
    that makes the time adjustment according to relative time, \emph{i.e.}, $f(C) := \{(a, t - t_{min}(C)) | (a, t) \in C \}$.
\end{enumerate}
This transformation of the event log allows us to compute average (and roughly analogously: median and mode) times at which an event occurs relative to the start time of the process.
We define a function $g_L: A \rightarrow \mathbb{R}$, \emph{i.e.}, for $a \in A$, we define $g(a) := avg(\{t |(a', t) \in \bigcup_{C \in L} C, a' = a \})$. Given the average occurrence time, we can then construct the timeline.
Note that the transformation is executed on a copy of the event log, \emph{i.e.}, we retain the initial log.





\subsection{Layout Strategy}
After calculating the average relative time points, the process model is discovered and aligned with the timeline. This involves creating a time axis representing the temporal distance between each node. By assuming a software-supported implementation using a graphical rendering and layout engine, the following steps are to be executed to create the timeline-based DFG:


\begin{enumerate}

    \item \textbf{Graph initialization.} Create a digraph object with the appropriate layout engine to show the hierarchical structure. This ensures that the graph's vertices are drawn in horizontal rows with the edges typically directed downwards (with the exception of exact or rounded-off equality of average occurrence times of several activities, where the edges run exactly horizontally).
    \item \label{two} \textbf{Label generation.} Round each timestamp in the mapped object to the nearest second, minute, month, and year, and format the values as human-readable \emph{strings}. Two examples of this are the time values \emph{0 days 00:12:00.540670} and \emph{128 days 22:46:09.230769230}, which are converted to \emph{12m} and \emph{4MO}, respectively.
    Each unique rounded-off value will be utilized as labels in nodes along the time axis. Furthermore, it is used in the edge length calculation. 
    These nodes represent time points at where events in the process model have occurred. 
    \item \textbf{Node generation.} Create the nodes of the time axis using graph visualization software and assign a unique identifier to each node. A record of the unique identifier is needed to later align the nodes of the process model with the nodes of the time axis.
    \item \textbf{Edge length generation.} For each subsequent node in time axis, we subtract the rounded off numerical value assigned to them during label generation.   
    \item \textbf{Edge generation.} Create the edges between the nodes of the time axis. This is accomplished by assigning a length to each edge based on the temporal distance between time axis nodes calculated using the time values created in step~\ref{two}.
    \item  \textbf{DFG generation.} Generate the DFG based on the initial (non-transformed) event log by applying a process discovery technique. The nodes are created for each activity and are assigned a unique identifier. The edges are assigned a label of frequency.
    \item \label{eight} \textbf{DFG-timeline mapping.} Map each activity in the DFG to a node on the timeline using their designated timestamps. Each node on the timeline will have one or more activities associated with it. Note that a node of the time axis may have more than one associated activity (1-to-many relationship). This information is maintained, as it is required for aligning the activities with their respective time axis node in the subsequent step.
    \item \textbf{Vertical alignment.} Finally, the information from the previous step is utilized to align the activities in the graph along the time axis according to their respective height on the canvas.
    This is achieved by creating subgraphs that are equal to the number of nodes on the time axis. Each subgraph includes the unique identifier of the time axis node. Additionally, it includes the identifiers referencing those activities that are supposed to be aligned with the current time axis node.
    
\end{enumerate}

\subsection{Prototypical Implementation}
The approach was implemented in a fork of the PM4Py library~\cite{DBLP:journals/corr/abs-1905-06169}\footnote{The code is available at \href{https://github.com/Har-leen-kaur/pm4py-timeline-axis}{https://github.com/Har-leen-kaur/pm4py-timeline-axis}. An example notebook can be found at \href{https://github.com/Har-leen-kaur/pm4py-timeline-axis/blob/release/6_timeline_axis_dfg.ipynb}{https://github.com/Har-leen-kaur/pm4py-timeline-axis/blob/release/6\_timeline\_axis\_dfg.ipynb}.}.


In the implementation, the program goes through the steps of our approach, utilizing the \emph{Graphviz}\footnote{https://graphviz.org/} to generate a script in the DOT language, Graphviz's abstract graph grammar.
The DOT script specifies what to include and where to position the nodes (and edges) of the graph.

\begin{figure} [h!]
    \centering
      \theverbbox \hspace{-30pt}
    \includegraphics[width=3.875cm]{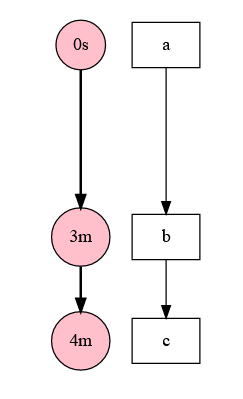}
    \caption{A simple example of a visualization result (right) after generating the DOT script (left) from directly-follows and time relations of three activities.}
    \label{fig:dotgraph}
\end{figure}

\emph{Example}: Let us introduce an example of the resulting DOT script. 
Assume we have an event log with four activities $a$, $b$, and $c$ and the directly-follows relations are $(a, b)$ and $(b, c)\}$. $a$ is the start activity. The relative time taken for each of the activities to execute is as follows: $a$: $0s$,  $b$: $3m$,  $c$: $4m$. Following the steps mentioned above will result in the graph shown in Figure~\ref{fig:dotgraph}. The DOT script from which the graph is rendered is shown next to the graph. 

\begin{verbbox}[\footnotesize]
digraph {
  layout=dot
  node [shape=rect <style details>]
  0 [label = "0s", <style details>]
  3 [label = "3m", <style details>]
  4 [label = "4m", <style details>]
  0 -> 3 [minlen=3 style=bold]
  3 -> 4 [minlen=1 style=bold]
  a [label= "a"]
  b [label= "b"]
  c [label= "c"]
  a -> b
  b -> c
  subgraph 0 {
    rank=same
    0 
    a
  }
  subgraph 3 {
    rank=same
    3
    b
  }
  subgraph 4 {
    rank=same
    4
    c
  }
} 
\end{verbbox}

\section{Evaluation}\label{eval}
In this section, we present an evaluation of the effectiveness of our approach. To this end, we compare the output of a classic DFG representation (in our case rendered by PM4Py) with the one from our novel approach. Our comparison focuses on the following questions. First, we count how many pairs of activities in the standard layout are positioned in a way that contradicts their temporal order. Technically, this is achieved by comparing the total preorders that emerge from the visual, top-to-bottom layout of the DFGs. Second, we discuss to which extent the temporal ordering provides a graph segmentation that is not visible with a standard layout. Finally, we investigate to what extent the distances along the time axis provide meaningful cues to the process analyst.

\subsection{Datasets used for comparison}

For this evaluation, we apply our implementation on three different event logs. First, we use two event logs from the BPI Challenge 2012\footnote{\url{https://doi.org/10.4121/uuid:3926db30-f712-4394-aebc-75976070e91f}} and 2017\footnote{\url{https://doi.org/10.4121/uuid:5f3067df-f10b-45da-b98b-86ae4c7a310b}}, which are based on real-life data of a loan application processes from a Dutch financial institute. The BPI Challenge 2012 event log comprises 262,200 events and 13,087 cases, and contains information for a loan and overdraft approvals process from submission to eventual resolution (approval, cancellation, or rejection).\footnote{\url{https://www.win.tue.nl/bpi/2012/challenge.html}} The event log from the BPI Challenge 2017 comprises 1,202,267 events and 31,509 cases and contains all applications filed through the institute's online system in 2016 and their subsequent events until February 1st, 2017, 15:11.\footnote{\url{https://www.win.tue.nl/bpi/2017/challenge.html}} We have a total order imposed by time. 

Apart from the first two event logs, we additionally use a proprietary event log of a sales process. The process instances were executed as part of a business-to-business software sales process of a medium-sized European company.
In this process, a \emph{lead} (customer contact) is generated and eventually converted into an \emph{opportunity} (lead with specific purchase scenario) that is then either lost/closed or turned into a paying customer.
The data cannot be shared for privacy-related ethics and compliance reasons; thus, only the resulting graphs are illustrated.
Furthermore, because of the complexity of the data, only a small sample (0.5\% of the event data) was considered.

\subsection{Comparison for Dataset 1} \label{sec:comparison1}

\begin{figure}[]%
    \centering
    \subfloat[\centering \label{result:bpic12-a} Directly follows graph with PM4Py standard layout]{{\includegraphics[width=0.45\textwidth]{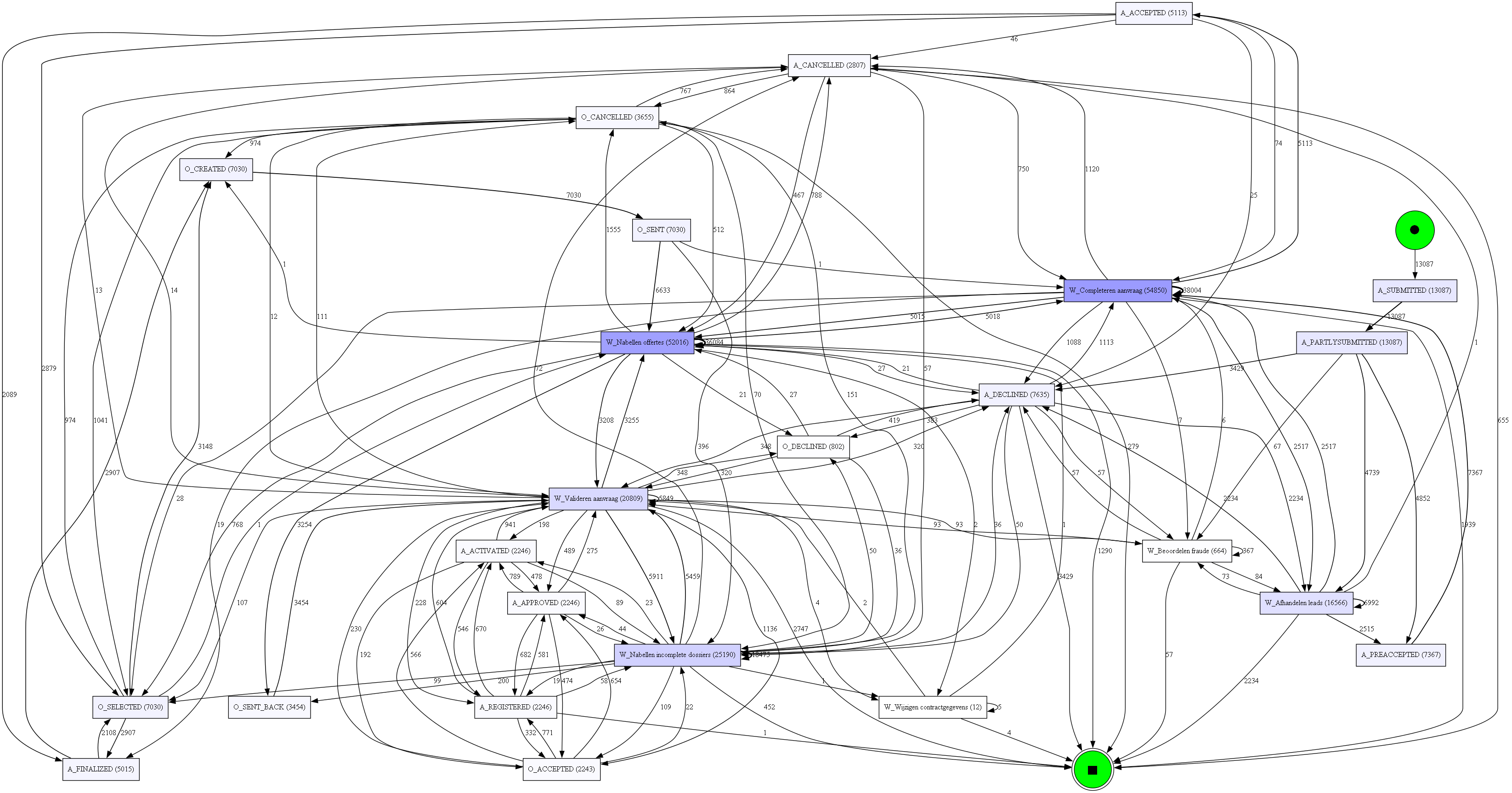} }}%
    \qquad
    \subfloat[\centering \label{result:bpic12-b} Timeline-based directly follows graph]{{\includegraphics[width=0.45\textwidth]{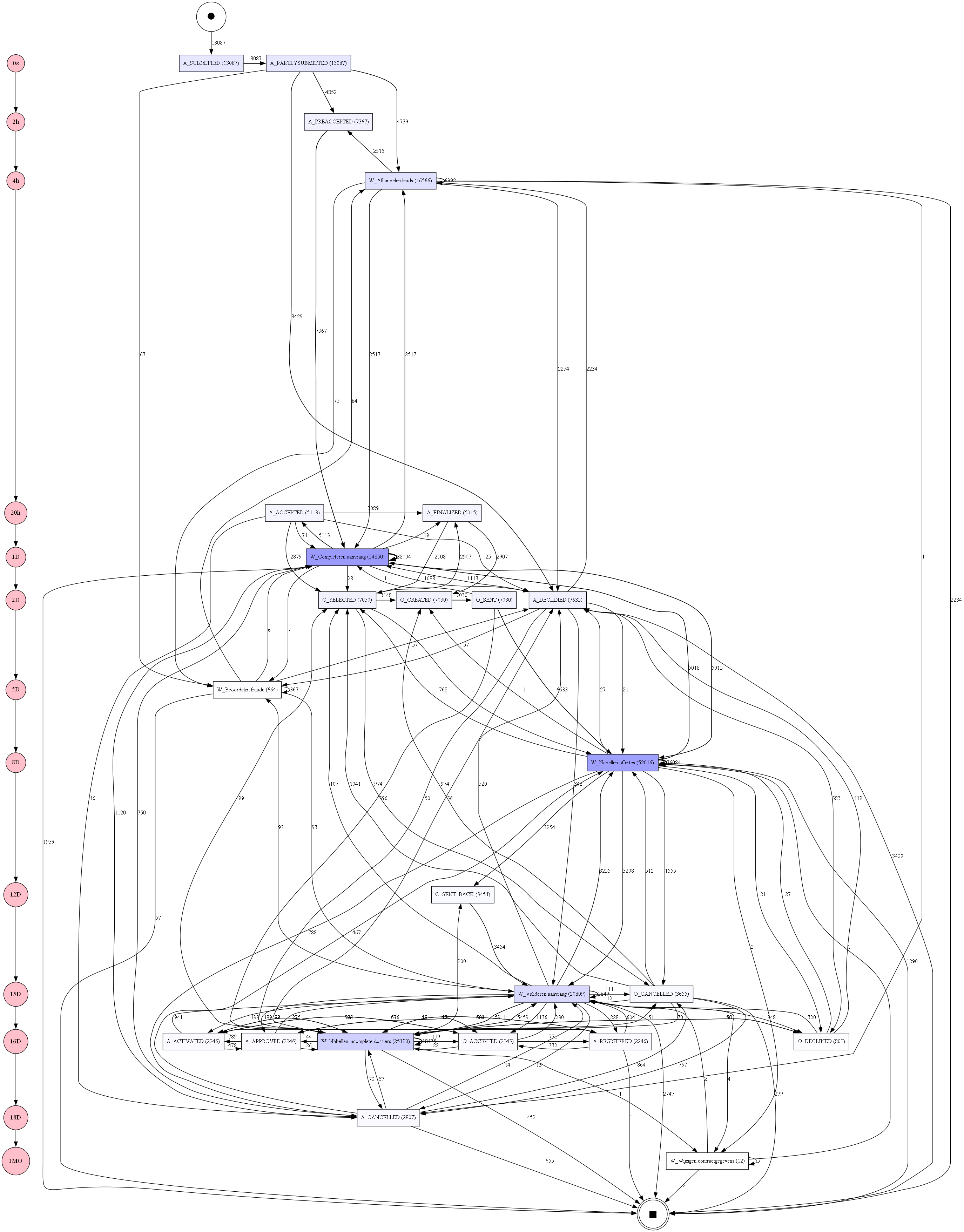} }}%
    \caption{BPI Challenge '12 event log: a simple DFG with standard layout and a DFG based on a timeline showing frequencies.}%
    \label{results:bpic12}%
\end{figure}

Figure \ref{results:bpic12} shows the resulting DFGs (standard and timeline-based layout) after applying our approach to the BPI Challenge 2012 event log. 

In the standard DFG, only four of the 24 activities (``W\_Completeren aanvraag, A\_Declined, W\_Nabellen incomplete dossiers, A\_REGISTERED'') are consistent with the temporal order as displayed in the timeline-based DFG.
In contrast, the timeline-based graph orders the activities according to their approximative relative temporal occurrence. This results in a more linear model and, presumably, a more logical impression of control-flow. An example of this is the start activity ``A\textunderscore Submitted'', which visually denotes the start of the process (see Figure~\ref{result:bpic12-b}) rather than appearing in the middle of the canvas (see Figure~\ref{result:bpic12-a}). Also, the end activities are now located toward the end of the process rather than visually dispersed — which is especially apparent for the activity ``A\textunderscore Cancelled''. Moreover, due to the hierarchical structure of the timeline-based graph, activities with similar approximate time points are vertically aligned, providing a sense of temporal concurrency, e.g., the row of activities at the time point \textit{16D} in Figure~\ref{result:bpic12-b}.

The timeline-based approach provides process analysts with additional temporal cues unavailable in the classical DFG layout. A noticeable improvement is the inclusion of the timeline itself, which allows for a holistic view of performance for the entire process — rather than for single process segments. As this results in a hierarchical process model with activities grouped into temporal segments, hence; the timeline-based graph also eases the finding of concurrent events or possible decision points. However, the simplified view of time also leads to further challenges. First, accurate time analysis is difficult because the activities align on an approximative logarithmic scale based on their average execution time. Second, the temporal concurrencies of activities that appear visually on the canvas, such as those at node \textit{16D}, make causal relations between these activities harder to read (cf., Figure~\ref{result:bpic12-b}). Lastly, some control-flow cannot be adequately interpreted, such as the edges that appear to go "backward" in time.

\subsection{Comparison for Dataset 2} \label{sec:comparison2}

\begin{figure}[ht!]%
    \centering
    \subfloat[\centering \label{result:bpic17-a} DFG with PM4Py standard layout]{{\includegraphics[width=0.45\textwidth]{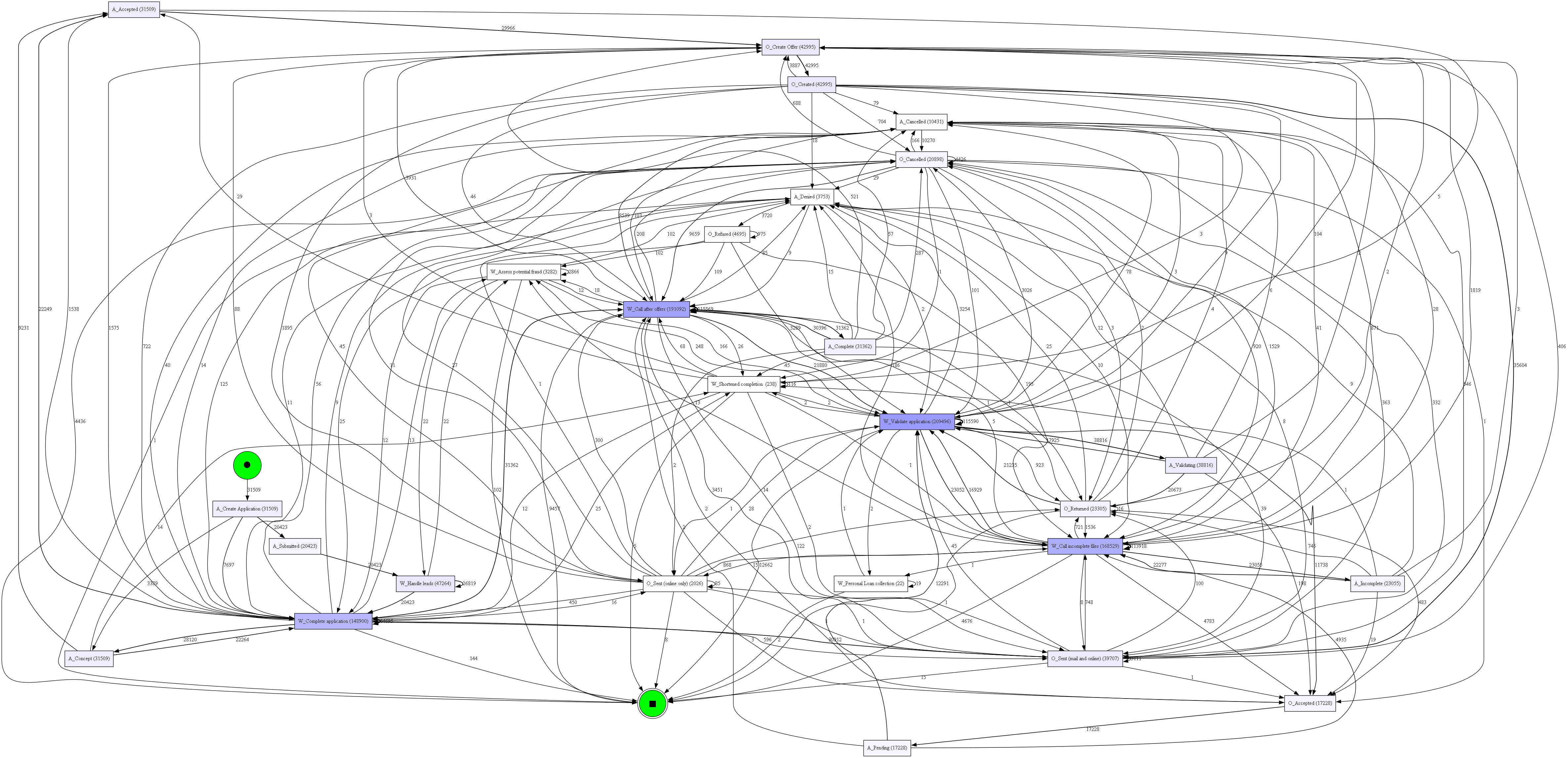} }}%
    \qquad
    \subfloat[\centering \label{result:bpic17-b} Timeline-based DFG]{{\includegraphics[width=0.45\textwidth]{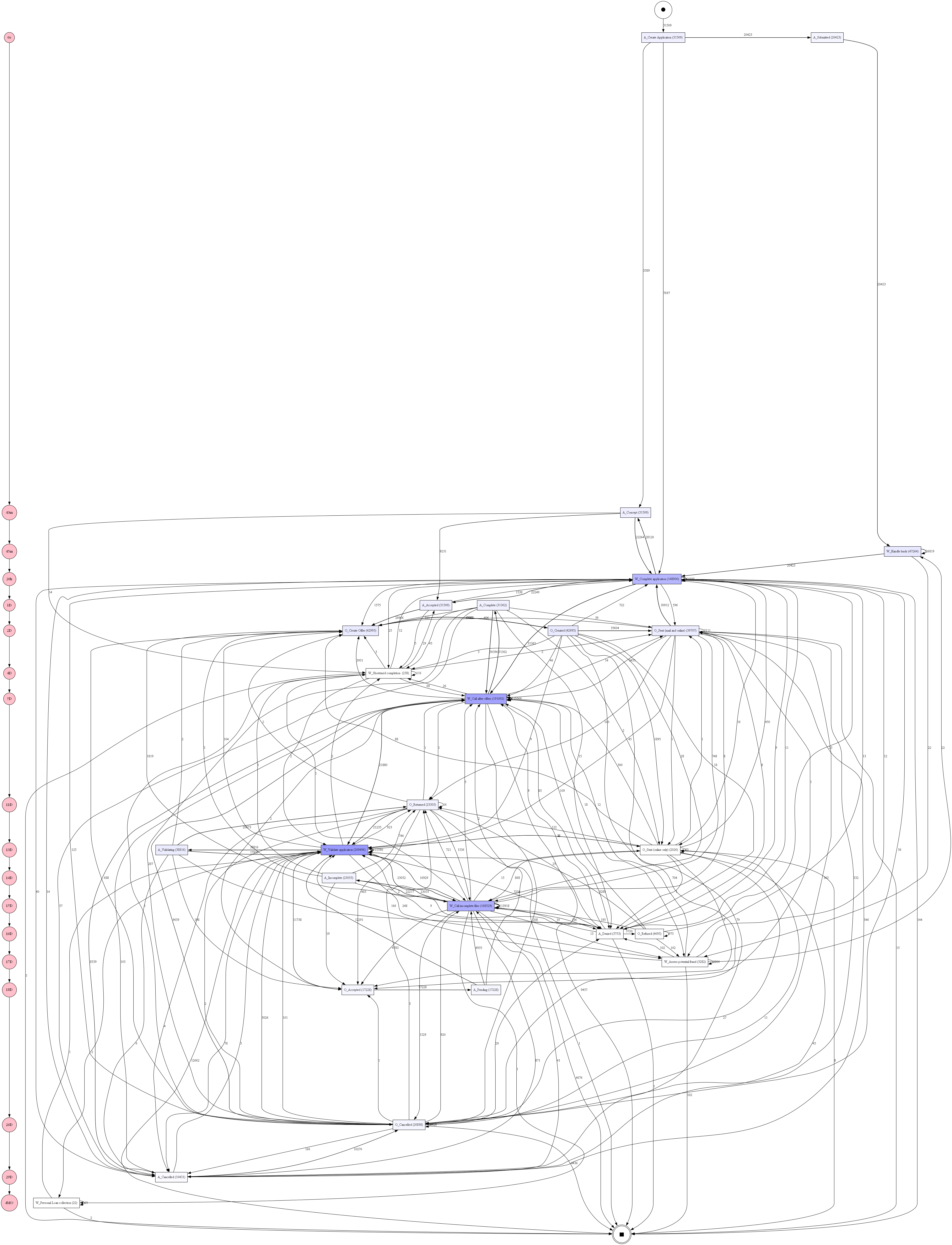} }}%
    \caption{BPI Challenge '17 event log: a simple DFG and DFG based on a timeline showing frequencies.}%
    \label{results:bpic17}%
\end{figure}
Figure \ref{results:bpic17} depicts the resulting DFGs after applying our approach to the BPI Challenge 2017 event log.

In the standard DFG, none of the 26 activities are consistent with the temporal order as depicted in the timeline-based DFG. This outcome might result from the low position of the starting activity ``A\_Create Application''. The following activities are accurately ordered visually after the starting event; yet their dependency on the start event leads to a contradiction of the temporal order (cf., Figure \ref{result:bpic17-a}).

As in the previous data set (BPIC 2012), the timeline-based DFG changes the structure of the process graph, however, with fewer apparent temporal segmentations of activities (see Figure \ref{result:bpic17-b}). Whereas the start activity ``A\_Create Application'' including the ``A\_Submitted'' are positioned at the top right corner of the timeline-based DFG, most activities follow a spaghetti-like pattern at the bottom half of the canvas.

The visual benefits and the limitations of the timeline-based DFG seen in the first comparison (section \ref{sec:comparison1}) are also prominent in this case. Yet, there are fewer rows of vertically aligned activities. Still, the spacing of time points suggests different process stages with approximate duration.
This cue is not visible in the standard DFG. For example, roughly four phases are visible: an application phase, a submission phase, a validation phase, and a cancellation phase. In addition, the timeline can be used to study the differences in process duration for different cases depending on their outcome. For example, a loan application that is later denied takes approximately 16 days, whereas a loan application (offer) later accepted takes approximately 18 days. An outlier activity is also apparent:  ``W\_Personal Loan collection'' extends the process duration by an estimated additional three months. None of these visual cues is evident in the standard DFG.

\subsection{Comparison for Dataset 3}

\begin{figure}[ht!]%
    \centering
    \subfloat[\centering \label{result:prop-a} DFG with PM4Py standard layout]{{\includegraphics[width=0.45\textwidth]{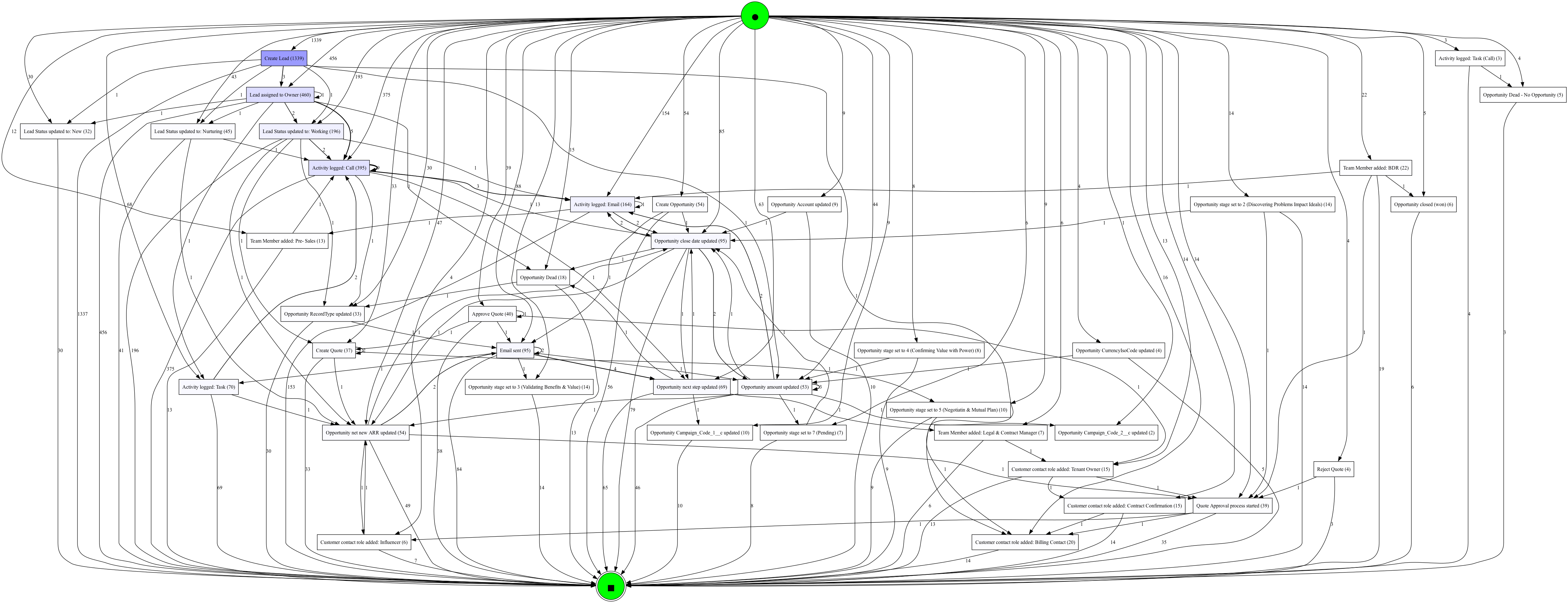} }}%
    \qquad
    \subfloat[\centering \label{result:prop-b} Timeline-based DFG]{{\includegraphics[width=0.45\textwidth]{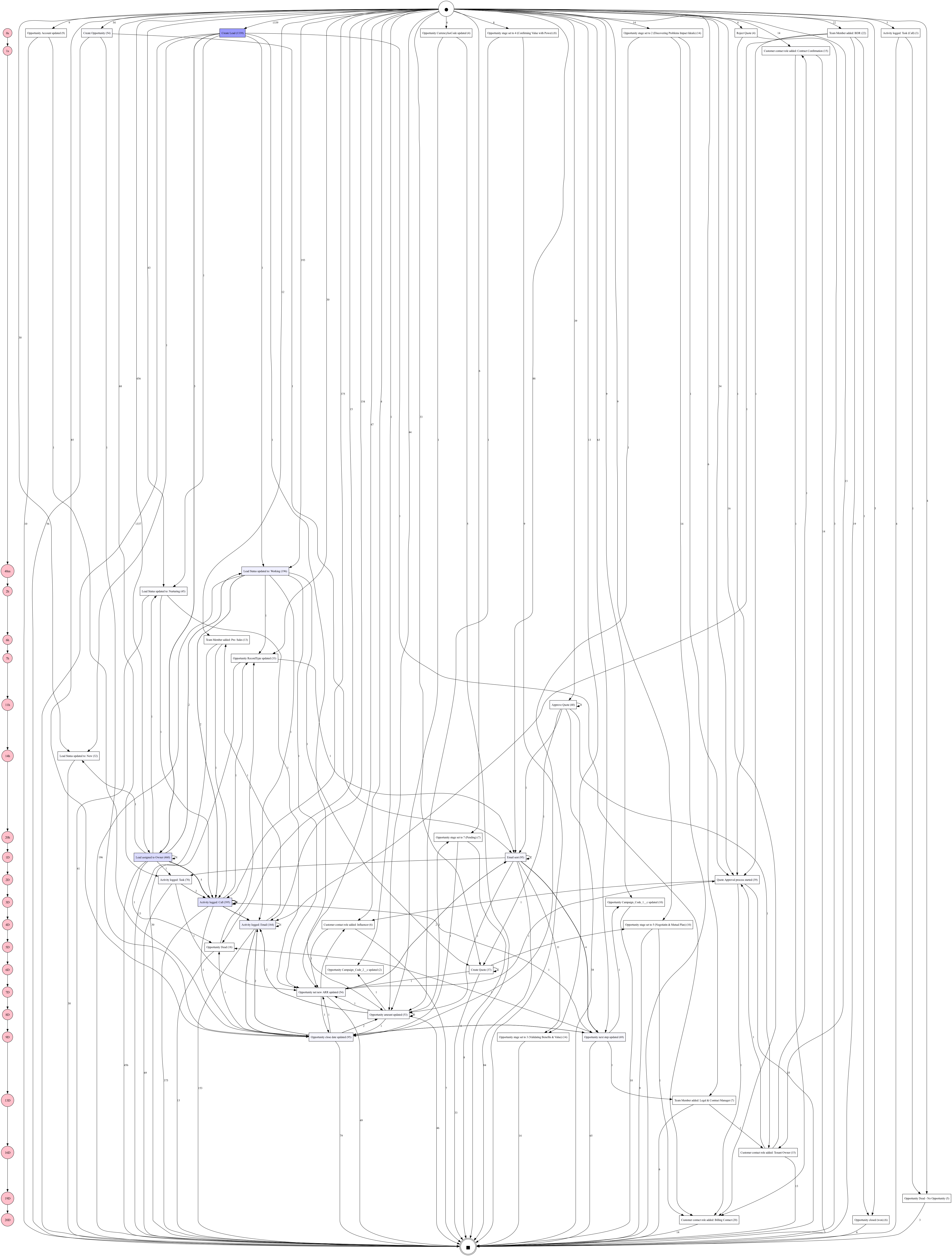} }}%
    \caption{Proprietary sales process dataset: a simple DFG and a DFG based on a timeline showing frequencies.}%
    \label{results:prop}%
\end{figure}

Figure \ref{results:prop} illustrates the resulting DFGs after applying our approach to the proprietary sales process data.

In the standard DFG, only six of the 24 activities (``Create Lead'', ``Activity logged: Task (Call)'', ``Team Member added: BDR'', ``Opportunity net new ARR updated'',
``Opportunity Campaign\_Code\_2\_\_c updated'' and ``Customer contact role added: Billing Contact'') are consistent with the temporal order displayed in the timeline-based DFG. Additionally, the starting and ending symbols of the standard DFG are positioned respectively at the top and the end of the canvas, giving the false illusion of a temporally ordered graph.
The timeline-based counterpart generates a more accurate depiction of the temporal order.

Similar to the previous comparison (\ref{sec:comparison2}), the timeline-based approach, in this case, visually reflects a notion of phases in the process model. In this process, a \emph{lead} (customer contact) is generated and eventually converted into an \emph{opportunity} (lead with specific purchase scenario) that is then either lost/closed or turned into a paying customer. Four phases are visually noticeable, such as a lead creation phase (first seconds), a lead setup phase (first hours), an opportunity nurturing phase (first week), and a closing phase (last two to three weeks). In this regard, the timeline does not only provide temporal cues but can reveal important details about the sales process and its critical stages.

Because the sales event log is highly complex, with many trace variations, this results in spaghetti-like graphs in both DFGs. However, the timeline-based DFG captures some of these variations indirectly through the vertical alignments of activities at their respective time points. This is especially evident for the different start activities.

\subsection{Discussion}
Based on the three comparisons, we can summarize the effectiveness of our approach as follows. 

First, our timeline-based approach effectively aligns a process model along a time axis, ordering the activities according to their respective estimated temporal occurrence. The standard DFG layout, at least as provided by PM4Py, is not a viable alternative in this regard because of its many contradictions to temporal order.

Second, the timeline-based DFG layout displays the process model in a hierarchical structure, thus providing a more intuitive visualization of processes. Starting activities are visualized at the top, whereas ending activities are positioned toward the bottom of the canvas. This is only sometimes the case in a standard DFG layout. Furthermore, activities that are approximately executed simultaneously are vertically aligned on the same row, forming temporal groups. These groups result in an additional depth of a process model, as they may indicate phases. However, the vertical aligning of activities also comes at the cost of a proper visual representation of the causal dependencies between these activities.

Finally, the timeline-based approach provides critical visual cues on process performance for analysts, cues that are not prominent or unavailable in the standard layout. The integrated time axis in the timeline-based DFG allows for a holistic view of performance, as it can be analyzed for the entire process rather than just on fragments. This supports the finding of bottlenecks and delays but also temporal outliers. In addition, the timeline gives indications of discrepancies in duration for various process paths. Nevertheless, the analytical capabilities of our approach are limited to approximations since it lacks a precise time description. There are also challenges to interpreting some visual depictions of control flow, such as edges that go in the opposite direction of the time axis.

\section{Conclusion}\label{conclusion}
In this paper, we addressed the challenge of providing insights into the temporal distances of activities of a process model. To that end, we developed a time axis proportional to the occurrences of events in a process. Our evaluation using three different event log datasets demonstrates that implementing a timeline in a process model is shown to be a viable approach for visualizing temporal data. The timeline axis provides an intuitive way for users to view temporal data. Further research is needed to evaluate the performance of the timeline axis, investigate its usefulness, as well as to explore other possible implementations of the timeline axis in process mining tools. 




\begin{thebibliography}{10}
\providecommand{\url}[1]{#1}
\csname url@samestyle\endcsname
\providecommand{\newblock}{\relax}
\providecommand{\bibinfo}[2]{#2}
\providecommand{\BIBentrySTDinterwordspacing}{\spaceskip=0pt\relax}
\providecommand{\BIBentryALTinterwordstretchfactor}{4}
\providecommand{\BIBentryALTinterwordspacing}{\spaceskip=\fontdimen2\font plus
\BIBentryALTinterwordstretchfactor\fontdimen3\font minus
  \fontdimen4\font\relax}
\providecommand{\BIBforeignlanguage}[2]{{%
\expandafter\ifx\csname l@#1\endcsname\relax
\typeout{** WARNING: IEEEtran.bst: No hyphenation pattern has been}%
\typeout{** loaded for the language `#1'. Using the pattern for}%
\typeout{** the default language instead.}%
\else
\language=\csname l@#1\endcsname
\fi
#2}}
\providecommand{\BIBdecl}{\relax}
\BIBdecl

\bibitem{DBLP:journals/is/WeerdtBVB12}
\BIBentryALTinterwordspacing
J.~D. Weerdt, M.~D. Backer, J.~Vanthienen, and B.~Baesens, ``A
  multi-dimensional quality assessment of state-of-the-art process discovery
  algorithms using real-life event logs,'' \emph{Inf. Syst.}, vol.~37, no.~7,
  pp. 654--676, 2012. [Online]. Available:
  \url{https://doi.org/10.1016/j.is.2012.02.004}
\BIBentrySTDinterwordspacing

\bibitem{polyvyanyy2020monotone}
A.~Polyvyanyy, A.~Solti, M.~Weidlich, C.~D. Ciccio, and J.~Mendling, ``Monotone
  precision and recall measures for comparing executions and specifications of
  dynamic systems,'' \emph{ACM Transactions on Software Engineering and
  Methodology (TOSEM)}, vol.~29, no.~3, pp. 1--41, 2020.

\bibitem{DBLP:journals/tkde/AugustoCDRMMMS19}
A.~Augusto, R.~Conforti, M.~Dumas, M.~L. Rosa, F.~M. Maggi, A.~Marrella,
  M.~Mecella, and A.~Soo, ``Automated discovery of process models from event
  logs: Review and benchmark,'' \emph{{IEEE} Trans. Knowl. Data Eng.}, vol.~31,
  no.~4, pp. 686--705, 2019.

\bibitem{DBLP:books/sp/DumasRMR18}
M.~Dumas, M.~L. Rosa, J.~Mendling, and H.~A. Reijers, \emph{Fundamentals of
  Business Process Management, Second Edition}.\hskip 1em plus 0.5em minus
  0.4em\relax Springer, 2018.

\bibitem{yeshchenko2022survey}
A.~Yeshchenko and J.~Mendling, ``A survey of approaches for event sequence
  analysis and visualization using the esevis framework,'' \emph{arXiv preprint
  arXiv:2202.07941}, 2022.

\bibitem{denisov2019predictive}
V.~Denisov, D.~Fahland, and W.~M. van~der Aalst, ``Predictive performance
  monitoring of material handling systems using the performance spectrum,'' in
  \emph{2019 International Conference on Process Mining (ICPM)}.\hskip 1em plus
  0.5em minus 0.4em\relax IEEE, 2019, pp. 137--144.

\bibitem{marey1875methode}
E.~Marey, ``La m{\'e}thode graphique dans les sciences exp{\'e}rimentales
  (suite),'' \emph{Physiologie Exp{\'e}rimentale. G. Masson}, 1875.

\bibitem{song2007supporting}
M.~Song and W.~M. van~der Aalst, ``Supporting process mining by showing events
  at a glance,'' in \emph{Proceedings of the 17th Annual Workshop on
  Information Technologies and Systems (WITS)}, 2007, pp. 139--145.

\bibitem{DBLP:series/hci/AignerMST11}
W.~Aigner, S.~Miksch, H.~Schumann, and C.~Tominski, \emph{Visualization of
  Time-Oriented Data}, ser. Human-Computer Interaction Series.\hskip 1em plus
  0.5em minus 0.4em\relax Springer, 2011.

\bibitem{mendling-nordsieck}
J.~Mendling, ``Business process modeling in the 1920s and 1930s as reflected in
  fritz nordsieck’s phd thesis,'' \emph{Enterprise Modelling and Information
  Systems Architectures - International Journal of Conceptual Modelling}, 2021.

\bibitem{DBLP:journals/corr/abs-2006-14291}
Y.~Guo, S.~Guo, Z.~Jin, S.~Kaul, D.~Gotz, and N.~Cao, ``A survey on visual
  analysis of event sequence data,'' \emph{IEEE Transactions on Visualization
  and Computer Graphics}, pp. 1--20, 2021.

\bibitem{DBLP:journals/tvcg/ChenXR18}
Y.~Chen, P.~Xu, and L.~Ren, ``Sequence synopsis: Optimize visual summary of
  temporal event data,'' \emph{{IEEE} Trans. Vis. Comput. Graph.}, vol.~24,
  no.~1, pp. 45--55, 2018.

\bibitem{Kwon2020DPVisVA}
B.~C. Kwon, V.~Anand, K.~Severson, S.~Ghosh, Z.~sun, B.~Frohnert, M.~Lundgren,
  and K.~Ng, ``Dpvis: Visual analytics with hidden markov models for disease
  progression pathways.'' \emph{IEEE transactions on visualization and computer
  graphics}, 2020.

\bibitem{DBLP:journals/tvcg/MonroeLLPS13}
M.~Monroe, R.~Lan, H.~Lee, C.~Plaisant, and B.~Shneiderman, ``Temporal event
  sequence simplification,'' \emph{{IEEE} Trans. Vis. Comput. Graph.}, vol.~19,
  no.~12, pp. 2227--2236, 2013.

\bibitem{DBLP:journals/cgf/LeiteGMGK20}
R.~A. Leite, T.~Gschwandtner, S.~Miksch, E.~Gstrein, and J.~Kuntner, ``{NEVA:}
  visual analytics to identify fraudulent networks,'' \emph{Comput. Graph.
  Forum}, vol.~39, no.~6, pp. 344--359, 2020.

\bibitem{DBLP:journals/cgf/DortmontEW19}
M.~A. M.~M. van Dortmont, S.~van~den Elzen, and J.~J. van Wijk,
  ``Chronocorrelator: Enriching events with time series,'' \emph{Comput. Graph.
  Forum}, vol.~38, no.~3, pp. 387--399, 2019.

\bibitem{DBLP:journals/tvcg/VrotsouJC09}
K.~Vrotsou, J.~Johansson, and M.~Cooper, ``Activitree: Interactive visual
  exploration of sequences in event-based data using graph similarity,''
  \emph{{IEEE} Trans. Vis. Comput. Graph.}, vol.~15, no.~6, pp. 945--952, 2009.

\bibitem{DBLP:journals/ivs/VrotsouYC14}
K.~Vrotsou, A.~Ynnerman, and M.~Cooper, ``Are we what we do? exploring group
  behaviour through user-defined event-sequence similarity,'' \emph{Inf. Vis.},
  vol.~13, no.~3, pp. 232--247, 2014.

\bibitem{DBLP:journals/cgf/RosenthalPMO13}
P.~Rosenthal, L.~Pfeiffer, N.~H. M{\"{u}}ller, and P.~Ohler, ``Visruption:
  Intuitive and efficient visualization of temporal airline disruption data,''
  \emph{Comput. Graph. Forum}, vol.~32, no.~3, pp. 81--90, 2013.

\bibitem{DBLP:journals/cgf/HanRDAS15}
Y.~Han, A.~Rozga, N.~Dimitrova, G.~D. Abowd, and J.~T. Stasko, ``Visual
  analysis of proximal temporal relationships of social and communicative
  behaviors,'' \emph{Comput. Graph. Forum}, vol.~34, no.~3, pp. 51--60, 2015.

\bibitem{DBLP:journals/tvcg/NguyenTAATZ19}
P.~H. Nguyen, C.~Turkay, G.~L. Andrienko, N.~V. Andrienko, O.~Thonnard, and
  J.~Zouaoui, ``Understanding user behaviour through action sequences: From the
  usual to the unusual,'' \emph{{IEEE} Trans. Vis. Comput. Graph.}, vol.~25,
  no.~9, pp. 2838--2852, 2019.

\bibitem{DBLP:journals/tvcg/VrotsouN19}
K.~Vrotsou and A.~Nordman, ``Exploratory visual sequence mining based on
  pattern-growth,'' \emph{{IEEE} Trans. Vis. Comput. Graph.}, vol.~25, no.~8,
  pp. 2597--2610, 2019.

\bibitem{DBLP:journals/tvcg/ZengWWWLEQ20}
H.~Zeng, X.~Wang, A.~Wu, Y.~Wang, Q.~Li, A.~Endert, and H.~Qu, ``Emoco: Visual
  analysis of emotion coherence in presentation videos,'' \emph{{IEEE} Trans.
  Vis. Comput. Graph.}, vol.~26, no.~1, pp. 927--937, 2020.

\bibitem{DBLP:journals/tvcg/LiuWWLL13}
S.~Liu, Y.~Wu, E.~Wei, M.~Liu, and Y.~Liu, ``Storyflow: Tracking the evolution
  of stories,'' \emph{{IEEE} Trans. Vis. Comput. Graph.}, vol.~19, no.~12, pp.
  2436--2445, 2013.

\bibitem{DBLP:journals/corr/abs-2008-09552}
T.~Baumgartl, M.~Petzold, M.~Wunderlich, M.~Hohn, D.~Archambault, M.~Lieser,
  A.~Dalpke, S.~Scheithauer, M.~Marschollek, V.~M. Eichel, N.~T. Mutters,
  H.~Consortium, and T.~V. Landesberger, ``In search of patient zero: Visual
  analytics of pathogen transmission pathways in hospitals,'' \emph{IEEE
  Transactions on Visualization and Computer Graphics}, vol.~27, no.~2, pp.
  711--721, 2021.

\bibitem{DBLP:journals/cgf/RedaTJLB11}
K.~Reda, C.~Tantipathananandh, A.~E. Johnson, J.~Leigh, and T.~Y.
  Berger{-}Wolf, ``Visualizing the evolution of community structures in dynamic
  social networks,'' \emph{Comput. Graph. Forum}, vol.~30, no.~3, pp.
  1061--1070, 2011.

\bibitem{DBLP:journals/tvcg/XuMR017}
P.~Xu, H.~Mei, L.~Ren, and W.~Chen, ``Vidx: Visual diagnostics of assembly line
  performance in smart factories,'' \emph{{IEEE} Trans. Vis. Comput. Graph.},
  vol.~23, no.~1, pp. 291--300, 2017.

\bibitem{DBLP:journals/tvcg/WuLYLW14}
Y.~Wu, S.~Liu, K.~Yan, M.~Liu, and F.~Wu, ``Opinionflow: Visual analysis of
  opinion diffusion on social media,'' \emph{{IEEE} Trans. Vis. Comput.
  Graph.}, vol.~20, no.~12, pp. 1763--1772, 2014.

\bibitem{DBLP:journals/cgf/SungHSCLW17}
C.~Sung, X.~Huang, Y.~Shen, F.~Cherng, W.~Lin, and H.~Wang, ``Exploring online
  learners' interactive dynamics by visually analyzing their time-anchored
  comments,'' \emph{Comput. Graph. Forum}, vol.~36, no.~7, pp. 145--155, 2017.

\bibitem{DBLP:journals/tvcg/FuldaBM16}
J.~Fulda, M.~Brehmer, and T.~Munzner, ``Timelinecurator: Interactive authoring
  of visual timelines from unstructured text,'' \emph{{IEEE} Trans. Vis.
  Comput. Graph.}, vol.~22, no.~1, pp. 300--309, 2016.

\bibitem{DBLP:journals/tvcg/JoHPKS14}
J.~Jo, J.~Huh, J.~Park, B.~H. Kim, and J.~Seo, ``Livegantt: Interactively
  visualizing a large manufacturing schedule,'' \emph{{IEEE} Trans. Vis.
  Comput. Graph.}, vol.~20, no.~12, pp. 2329--2338, 2014.

\bibitem{DBLP:journals/cgf/LiuKDGHW17}
Z.~Liu, B.~Kerr, M.~Dontcheva, J.~Grover, M.~Hoffman, and A.~Wilson,
  ``Coreflow: Extracting and visualizing branching patterns from event
  sequences,'' \emph{Comput. Graph. Forum}, vol.~36, no.~3, pp. 527--538, 2017.

\bibitem{DBLP:journals/is/BoseA12}
R.~P. J.~C. Bose and W.~M.~P. van~der Aalst, ``Process diagnostics using trace
  alignment: Opportunities, issues, and challenges,'' \emph{Inf. Syst.},
  vol.~37, no.~2, pp. 117--141, 2012.

\bibitem{DBLP:journals/dss/LeoniAAH12}
M.~de~Leoni, M.~Adams, W.~M.~P. van~der Aalst, and A.~H.~M. ter Hofstede,
  ``Visual support for work assignment in process-aware information systems:
  Framework formalisation and implementation,'' \emph{Decis. Support Syst.},
  vol.~54, no.~1, pp. 345--361, 2012.

\bibitem{DBLP:journals/is/LowAHWW17}
W.~Z. Low, W.~M.~P. van~der Aalst, A.~H.~M. ter Hofstede, M.~T. Wynn, and J.~D.
  Weerdt, ``Change visualisation: Analysing the resource and timing differences
  between two event logs,'' \emph{Inf. Syst.}, vol.~65, pp. 106--123, 2017.

\bibitem{DBLP:journals/is/RichterS19}
F.~Richter and T.~Seidl, ``Looking into the {TESSERACT:} time-drifts in event
  streams using series of evolving rolling averages of completion times,''
  \emph{Inf. Syst.}, vol.~84, pp. 265--282, 2019.

\bibitem{DBLP:journals/dss/SuriadiOAH15}
S.~Suriadi, C.~Ouyang, W.~M.~P. van~der Aalst, and A.~H.~M. ter Hofstede,
  ``Event interval analysis: Why do processes take time?'' \emph{Decis. Support
  Syst.}, vol.~79, pp. 77--98, 2015.

\bibitem{DBLP:journals/dss/SuriadiWXAH17}
S.~Suriadi, M.~T. Wynn, J.~Xu, W.~M.~P. van~der Aalst, and A.~H.~M. ter
  Hofstede, ``Discovering work prioritisation patterns from event logs,''
  \emph{Decis. Support Syst.}, vol. 100, pp. 77--92, 2017.

\bibitem{DBLP:journals/corr/abs-1905-06169}
\BIBentryALTinterwordspacing
A.~Berti, S.~J. van Zelst, and W.~M.~P. van~der Aalst, ``Process mining for
  python (pm4py): Bridging the gap between process- and data science,''
  \emph{CoRR}, vol. abs/1905.06169, 2019. [Online]. Available:
  \url{http://arxiv.org/abs/1905.06169}
\BIBentrySTDinterwordspacing

\end{thebibliography}
\end{document}